# Deterministic X-ray Bragg coherent diffraction imaging as a seed for subsequent iterative reconstruction

Konstantin M. Pavlov[1,2,3], Kaye S. Morgan[2,4,5], Vasily I. Punegov[6] and David M. Paganin[2]

[1] School of Physical and Chemical Sciences, University of Canterbury, Christchurch, New Zealand
[2] School of Physics and Astronomy, Monash University, Clayton, Australia
[3] School of Science and Technology, University of New England, Armidale, Australia
[4] Institute of Advanced Studies, Technical University of Munich, Garching, Germany
[5] Chair of Biomedical Physics, Department of Physics, Technical University of Munich, Garching, Germany
[6] Komi Research Center, Ural Division, Russian Academy of Sciences, Syktyvkar, Russian Federation

E-mail: konstantin.pavlov@canterbury.ac.nz

**Abstract**

Coherent diffractive imaging (CDI), using both X-rays and electrons, has made extremely rapid progress over the past two decades. The associated reconstruction algorithms are typically iterative, and seeded with a crude first estimate. A deterministic method for Bragg Coherent Diffraction Imaging (Pavlov *et al.*, *Sci. Rep.* **7**, 1132 (2017)) is used as a more refined starting point for a shrink-wrap iterative reconstruction procedure. The appropriate comparison with the autocorrelation function as a starting point is performed. Real-space and Fourier-space error metrics are used to analyse the convergence of the reconstruction procedure for noisy and noise-free simulated data. Our results suggest that the use of deterministic-CDI reconstructions, as a seed for subsequent iterative-CDI refinement, may boost the speed and degree of convergence compared to the cruder seeds that are currently commonly used. We also highlight the utility of monitoring multiple error metrics in the context of iterative refinement.

## 1. Introduction

The "phase problem" for propagating complex scalar fields seeks to reconstruct both their phase and amplitude given measurements of wave-field modulus [1]. Such data may be directly obtained using experimental measurements of field intensity or probability density.

Phase retrieval has a rich history throughout many domains of optical and quantum physics, dating back at least as far as Wolfgang Pauli's famous question regarding the possibility of reconstructing a complex scalar wave-function given knowledge of the modulus of both its real-space and momentum-space wave-functions [2]. Accordingly, phase retrieval methodologies have been applied in many imaging-related fields including visible-light optics [3], X-ray optics [4-7], electron optics [8] and neutron optics [9]. While linear optics is typically considered, phase retrieval for non-linear fields (such as those obeying the non-linear Schrödinger equation) has also been studied [10]. The above rich variety of fields is accompanied by a variety of approaches to phase retrieval. These include but are not limited to interferometry [11]), holography (inline holography [12], off-axis holography [13-15], Fourier holography [16,17] *etc.*), through-focal series techniques [18], various means for the inversion of far field scattering data [5,19,20], ptychographic methods [21], and deliberate introduction of aberrations [22].

We restrict usage of the term "phase retrieval" to means of phase recovery that are not explicitly based on interferometry. Two features are common to many methods of phase retrieval. (i) Constructive use is made of the differential equation governing the evolution of the field, which couples the measured intensity and to-be-recovered phase and thereby permits one to pose the inverse problem of recovering the latter from the former. (ii) Use of relevant *a priori* knowledge is often crucial.

The phase-retrieval problem, of recovering phase information from a measurement of wave-field moduli, is an example of a so-called inverse problem [23]. A specified phase-retrieval scenario is "well-posed in the sense of Hadamard" if it satisfies the criteria of (i) existence of at least one solution, (ii) uniqueness of the solution modulo acceptable ambiguities such as meaningless global phase factors and transverse displacement of the object being reconstructed, and (iii) stability of the solution with respect to imperfections in the input intensity data (see *e.g.* p. 221 in [24]).

In deterministic approaches to phase retrieval [25], these three criteria for well-posedness may be explicitly addressed. This has the advantage of conceptual clarity and rigour, balanced against the negatives that (i) it severely restricts the scope of phase-retrieval problems that may be addressed; (ii) reconstruction errors can result from a realistic sample's deviation from the strong assumptions often needed to develop a deterministic solution.

A complementary strategy adopts iterative approaches to solving the inverse problem of phase retrieval [26,3]. Here, one typically sets up an error metric which quantifies the degree of mismatch between the data implied by a given candidate reconstruction (of the complex wave-field, or of a given object which has resulted in a measured wave-field). One seeks to minimise this error metric, subject to suitable constraints (such as the finite domain occupied by the object, atomicity and/or positivity of the object, *etc.*) and other relevant *a priori* knowledge. The approach pioneered by Gerchberg & Saxton [26] and Fienup [3], together with its successors (*e.g.* [27]), has been particularly successful. Such iterative approaches to phase retrieval have the advantage that they can be practically applied to a much broader class of problem than is amenable to deterministic approaches, while having the drawback that they can lack the conceptual clarity and rigour that deterministic methods provide. This drawback may be problematic, for example, when an iterative phase-retrieval algorithm is trapped in a non-global local minimum of the error metric, making it unclear whether the stagnated solution is indeed acceptably close to the correct solution.

Iterative and deterministic approaches to the inverse problem are not necessarily mutually exclusive. Deterministic phase-retrieval methods can be used to give a good first estimate to the solution to a specified phase problem, which can then be iteratively refined into a better solution. The key idea is that the deterministic method locates a point in the solution space that is sufficiently close to the global error-metric minimum corresponding to the true solution, thereby aiding both the rapidity and the correctness of the iterative-method convergence to a better solution to the particular phase problem.

We focus attention on "coherent diffractive imaging" (CDI) [28]. This relates to phase retrieval for non-crystalline (or imperfectly crystalline) samples using far-field optical scattering data. CDI is a non-destructive technique enabling nano-resolution imaging, particularly using X-rays and electrons, whose success has been demonstrated in a number of applications [29].

Until recently, most CDI reconstruction techniques available were iterative. An exception is given by methods related to Fourier holography [30], about which more will be said later. Contemporary iterative approaches to CDI have enjoyed an impressive chain of successes, with the associated iterative phase-retrieval methods having achieved a high level of accuracy and robustness.

Nevertheless, the previously-described issues intrinsic to iterative approaches are not entirely eliminated. Indicative is the following statement from a recent review: "The presence of noise and limited prior knowledge (loose constraints) increases the number of solutions within the noise level and constraints. Confidence that the recovered image is the correct and unique one can be obtained by repeating the phase-retrieval process using several random starts." [31]. Usually, these iterative reconstruction techniques use as a starting guess in real space an autocorrelation of the object function, obtained as the inverse Fourier transform of the far-field diffraction pattern (*e.g.* [27]) or a random set of parameters (*e.g.* the guided hybrid-input-output (HIO) method [32]).

We explore deterministic phase-retrieval seeding of subsequent iterative-method refinement in the problem of Bragg-CDI phase retrieval. Bragg CDI is a variant of CDI applied to small imperfect crystals, using 3D far-field diffracted-intensity measurements in the vicinity of a Bragg peak as data from which one seeks to reconstruct both the shape and strain-field distribution data in the crystal [33-35]. Typical crystal dimensions are on the order of tens of nanometres through to several microns. We are particularly interested in investigating whether an iterative technique can help to remove or reduce the above-mentioned reconstruction errors, produced by deterministic approaches, by using the reconstruction results of the deterministic method as a starting guess instead of *e.g.*, an autocorrelation-based function.

The particular technique we consider here is a new deterministic method for 3D Bragg CDI, which allows a non-destructive reconstruction of chemical composition and strain in facetted crystalline materials [36]. This builds upon a deterministic 2D reconstruction approach [30], which may be considered as a form of Fourier holography [16,17], and which was later extended and further developed by several groups (see the recent review on deterministic CDI by Allen *et al.* [25], together with references therein). Deviations from the specified crystalline shapes or quality produce errors in the obtained reconstructions as shown in [36]. In the present paper we show via X-ray simulations that such artefacts in the deterministic Bragg-CDI reconstruction may be improved via subsequent iterative refinement. Our key focus however is on the more interesting finding that both the rate and quality of the convergence of iterative Bragg-CDI may be improved by seeding the iterative Bragg-CDI algorithm with a deterministic-CDI reconstruction, rather than merely seeding it with auto-correlation-based or random initial guesses. Thus our simulations are consistent with the conclusions (i) that deterministic-Bragg-CDI errors are reduced by subsequent iterative Bragg-CDI refinement, and (ii) that the errors of iterative Bragg CDI are reduced by deterministic-Bragg-CDI seeding. Note that we are primarily interested in giving an example of the use of deterministic phase-retrieval methods to seed iterative phase-retrieval methods for Bragg CDI and hence improve both the rate and quality of convergence of the latter, rather than exploring the absolute state of the art in either iterative or deterministic methods for approaching such a phase problem.

## 2. Methods

Consider a plane monochromatic X-ray wave with sigma polarisation and unit intensity, which illuminates a small deformed crystal. The angle between the wave vector $\mathbf{k}$ of the incident wave and the X-axis is $\theta_1 = \theta_B + \Delta\theta_1$ (see Fig.1), where $\theta_B$ is the Bragg angle for a symmetrical (00L) type reflection, $\Delta\theta_1$ is the angular deviation and the Cartesian axes (X,Y,Z) are as defined in Fig. 1. The scattered wave is registered in the direction of the wave vector $\mathbf{k}'$. The angle between $\mathbf{k}'$ and the X-axis is $\theta_2 = \theta_B + \Delta\theta_2$ (see Fig.1), where $\Delta\theta_2$ is the angular deviation. The incident wave vector $\mathbf{k}$ and the average scattered wave vector $\mathbf{k}'$ lie in the diffraction plane XOZ and $|\mathbf{k}| = |\mathbf{k}'| = k = 2\pi/\lambda$, where $\lambda$ is the wavelength in vacuum. The plane of detector D is perpendicular to the vector $\mathbf{k}'$. The scattering vector, $\mathbf{Q}$, is defined as $\mathbf{Q} = \mathbf{k}' - \mathbf{k}$.

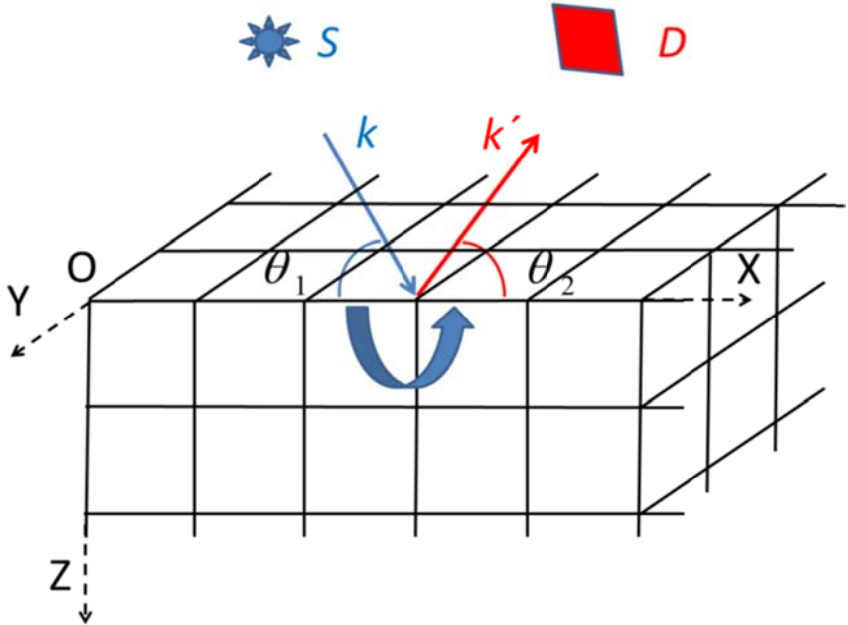

**Figure 1**. Sigma-polarised monochromatic X-rays from a source S impinge upon a cubic crystalline sample, for which only one corner is shown. Three-dimensional diffraction data is measured by a detector D which records diffracted intensities over a three-dimensional range of scattering vectors, centred about a particular scattering direction, specified by angles $\theta_1$ and $\theta_2$. Such diffracted intensity measurements comprise the data which may then be input into our reconstruction method for Bragg CDI [36], to give both the structure factor and the displacement field of the crystal.

We consider two-beam kinematical diffraction in a coplanar geometry, where the XOY plane is the top surface of the crystal. For simplicity, the crystalline structure is assumed to have cubic symmetry with lattice constant $a$ and parallelepiped shape with dimensions $L_x=N_x \cdot a$, $L_y=N_y \cdot a$ and $L_z=N_z \cdot a$, where the z-direction is vertical, XOY is the horizontal plane (cf. Fig.1) and $N_{x,y,z}$ are integers. For simplicity, we set $L_x=L_y=L_z$, i.e., $N_x=N_y=N_z$. To avoid any aliasing problems and allow a successful iterative reconstruction procedure we use $4\times4\times4=64$ oversampling ratio as defined in [37]. In our previous paper [36], we introduced a reconstruction method based on an assumption that at least half of the crystal is ideal, i.e., deformation and defect free. In this paper we assume that there are three spherical inclusions (structural defects) in the upper half of the crystal, as in [36]. We also assume existence of a curved deformation field in the entire crystal (see Fig. 2a) according to the following functional form (see [36]): $e^{-i\mathbf{h}\cdot\mathbf{u}(\mathbf{r})} = e^{i\gamma\left[(x-L_x/2)^2+(y-L_y/2)^2\right]\cdot[1-z/L_z]}$ for $z\in[0,L_z]$. This parabolic displacement may be associated [38] with the incorporation of the "non-ideal" part of the crystal to an "ideal" part of the crystal. This means that the deterministic Bragg CDI (BCDI) reconstruction algorithm, outlined in [36], will inevitably produce reconstruction errors, which will be then partly compensated by the subsequent application of an iterative refinement procedure. The constant used in describing the deformation field, $\gamma$, is inversely proportional to the radius of curvature of the deformation field. In our simulations, we have chosen the values of the coefficient $\gamma$ to yield a maximum phase shift of $0.25\pi$ at the edges of the crystal, so that displacements in the bottom half of the crystal are relatively small. Therefore, the deformation field in the bottom part can be considered as a small perturbation to the ideal model (i.e., a deformation-free bottom half of the crystal volume), as required for the deterministic Bragg CDI reconstruction algorithm [36]. The choice of maximum phase shift of $0.25\pi$ simplifies the reconstruction procedure as it does not require unwrapping of the reconstructed phase.

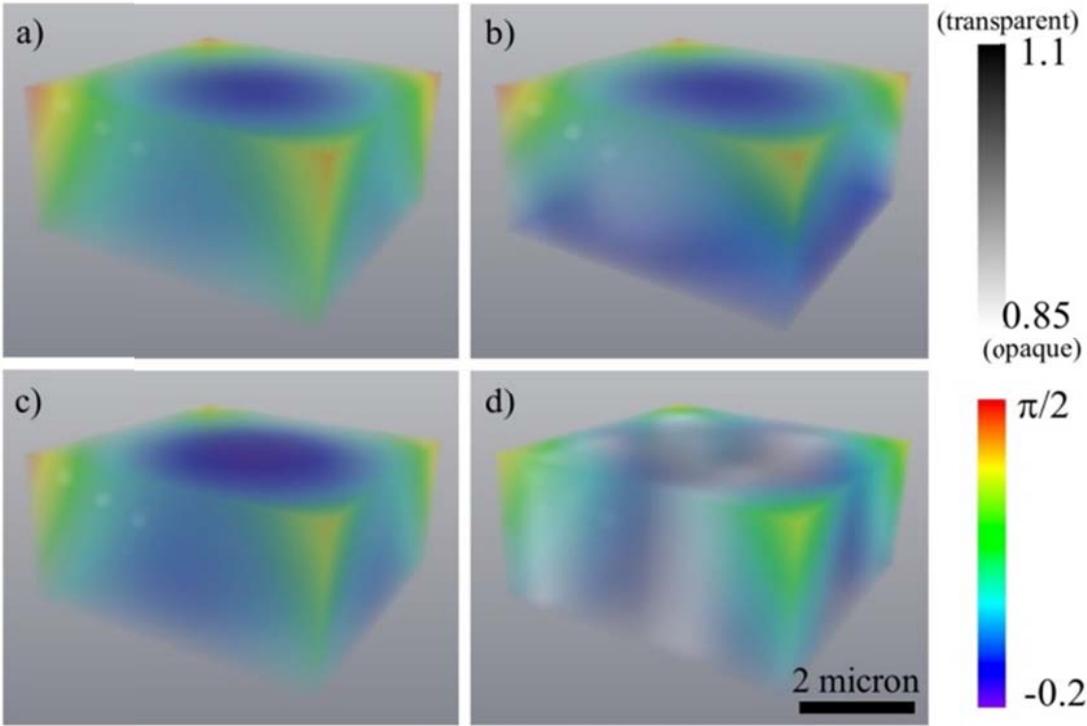

**Figure 2.** Visualisation of the phase (colour) and amplitude (greyscale) for a) the simulated crystal, b) the crystal reconstructed using the deterministic BCDI reconstruction [36], c) the crystal reconstructed using the shrink-wrap iterative procedure [27], with the BCDI reconstruction as a starting point and d) the crystal reconstructed using the shrink-wrap iterative procedure, with the autocorrelation function as a starting point.

The "target" function in our reconstructions will be $\beta_h(\mathbf{r})\exp(-i\mathbf{h}\cdot\mathbf{u}(\mathbf{r}))$ (cf. [36]), where $\beta_h(\mathbf{r}) = \chi_h(\mathbf{r})/\chi_h^{id}(\mathbf{r})$, $\chi_h(\mathbf{r})$ is the polarizability of a non-ideal crystal, $\chi_h^{id}(\mathbf{r})$ is the polarizability of an ideal crystal, $\mathbf{h}=\mathbf{Q}-\mathbf{q}$ is the reciprocal lattice vector for the chosen 00L reflection, $\mathbf{Q}$ is the scattering vector, and $\mathbf{u}(\mathbf{r})$ is the displacement vector field. For simplicity, we can neglect the imaginary component (as is usually done in BCDI, see *e.g.* [39]) of the function $\beta_h(\mathbf{r})$ because the imaginary parts of the functions $\chi_h(\mathbf{r})$ and $\chi_h^{id}(\mathbf{r})$ are typically significantly smaller than their corresponding real parts.

In this paper, we place weak inclusions in the upper half of the crystal, each filled with a material having a smaller structure factor than the bulk (minimum $\beta_h = 0.9$). The remainder of the simulated crystalline structure (*i.e.* all but the spherical inclusions) has $\beta_h = 1$ (see Fig. 2a showing $\beta_h(\mathbf{r})$ and $phase(\mathbf{r}) = -\mathbf{h}\cdot\mathbf{u}(\mathbf{r})$). The cropped data array of far-field intensity used in the reconstructions results in a voxel resolution in real space of 80×80×80 nm$^3$, which is comparable to the resolution demonstrated in [40]. However, it should be noted that a better resolution is reported in more recent literature (*e.g.* [41]). Application of the proposed reconstruction technique for a smaller voxel size is straightforward.

The expression for the simulated far-field intensity (without noise) is given in [36]:

$$I_{kin}/C = \left\{ N_x N_y N_z \frac{\sin(q_x a N_x/2)}{q_x a N_x/2} \frac{\sin(q_y a N_y/2)}{q_y a N_y/2} \frac{\sin(q_z a N_z/2)}{q_z a N_z/2} \right\}^2$$
$$+ N_x N_y N_z \frac{\sin(q_x a N_x/2)}{q_x a N_x/2} \frac{\sin(q_y a N_y/2)}{q_y a N_y/2} \frac{\sin(q_z a N_z/2)}{q_z a N_z/2}$$
$$\times e^{iq_x a(N_x-1)/2} e^{iq_y a(N_y-1)/2} e^{iq_z a(N_z-1)/2} Z(\mathbf{q}) \sum_{p=1}^{N} e^{-i\mathbf{q}\cdot\mathbf{R}_p} \left[ \beta_h^p e^{-i\mathbf{h}\cdot\mathbf{u}_p} - 1 \right]$$
$$+ N_x N_y N_z \frac{\sin(q_x a N_x/2)}{q_x a N_x/2} \frac{\sin(q_y a N_y/2)}{q_y a N_y/2} \frac{\sin(q_z a N_z/2)}{q_z a N_z/2} \qquad (1)$$
$$\times e^{-iq_x a(N_x-1)/2} e^{-iq_y a(N_y-1)/2} e^{-iq_z a(N_z-1)/2} Z^*(\mathbf{q}) \sum_{p=1}^{N} e^{i\mathbf{q}\cdot\mathbf{R}_p} \left[ \left(\beta_h^p\right)^* e^{i\mathbf{h}\cdot\mathbf{u}_p} - 1 \right]$$
$$+ \left| Z(\mathbf{q}) \sum_{p=1}^{N} e^{-i\mathbf{q}\cdot\mathbf{R}_p} \left[ \beta_h^p e^{-i\mathbf{h}\cdot\mathbf{u}_p} - 1 \right] \right|^2.$$

Here, $\mathbf{u}_p = \mathbf{u}(\mathbf{r})\cdot\delta(\mathbf{r}-\mathbf{R}_p)$, $\beta_h^p = \beta_h(\mathbf{r})\cdot\delta(\mathbf{r}-\mathbf{R}_p)$, $\mathbf{R}_p$ defines the position of the *p*-th cell in an ideal 3D periodic lattice, $Z(\mathbf{q}) = \sum_{l=1}^{R} \exp(-i\mathbf{q}\mathbf{r}_l)$ is an interference function, $\mathbf{r}_l$ defines the position of elementary cells within the *p*-th cell, *C* is a constant and $\delta(\mathbf{r})$ is the Dirac delta function.

If at least one half of the crystal is an ideal reference part (the bottom half in our case), we can obtain a closed-form solution to the BCDI inverse problem of reconstructing $\beta_h(\mathbf{r})\exp(-i\mathbf{h}\cdot\mathbf{u}(\mathbf{r}))$ (comp. equation (11) in [36]) by introducing the auxiliary function:

$$U(\mathbf{R}=(x,y,z))$$
$$= C\frac{1}{(2\pi)^{3/2}} \iiint q_x q_y q_z \hat{I}_{kin} e^{i(q_x x + q_y y + q_z z)} dq_x dq_y dq_z. \qquad (2)$$

Here $\hat{I}_{kin}$ is the intensity $I_{kin}$ with added Poisson noise. In our simulations, we consider two cases of noise applied to the simulated data, namely, noise free and with a maximum intensity of $10^{11}$ photons per voxel (at **q**=0). As we want to compare the effectiveness of deterministic and iterative reconstruction procedures in this paper, we have not excluded the brightest voxel, corresponding to the origin in Fourier space, from the noise adding procedure as was done in [36]. For real experimental data, $\hat{I}_{kin}$ is proportional to the registered intensity.

As shown in detail in [36], the auxiliary function *U(x,y,z)* reduces to

$$U(\mathbf{R}=(x,y,z)) = A + \sum_{j=1}^{8}(B_j + C_j) + D. \qquad (3)$$

Here, $A$ is a term proportional to the shape function of the sample, the $B_j$ and $C_j$ terms are 16 spatially-translated independent reconstructions of the unknown complex field $\beta_h^{rec}(\mathbf{r})\exp(i \cdot phase^{rec}(\mathbf{r})) = \beta_h^{rec}(\mathbf{r})\exp(-i\mathbf{h}\cdot\mathbf{u}^{rec}(\mathbf{r}))$ or its complex conjugate, and $D$ is related to the derivative of the cross-correlation of the object.

The closed-form solution to the BCDI inverse problem, obtained using equation (2), is exact if a) there is no noise in $\hat{I}_{kin}$ and b) the reference part is indeed a perfect undeformed crystal. If either of these conditions is not fulfilled, the reconstruction of $\beta_h(\mathbf{r})\exp(-i\mathbf{h}\cdot\mathbf{u}(\mathbf{r}))$ will contain some errors. The stronger the deviations from the ideal reconstruction conditions, the greater these errors will typically become.

To reduce the reconstruction errors associated with our deterministic BCDI algorithm, we employ the widely-used shrink-wrap iterative reconstruction algorithm [27]. The shrink-wrap algorithm, being very well known, will not be described here. In the original paper [27], it was suggested that one could use the autocorrelation function as a starting point for this iterative reconstruction algorithm. Such a choice has been commonly employed, in a large number of successful CDI reconstructions. However, as an alternative that is explored in the present paper, one can choose a BCDI reconstruction result as a starting point for subsequent iterative refinement.

## 3. Modelling, Results and Discussion

Now we apply our deterministic BCDI reconstruction (see equation (2)) to the simulated X-ray intensity (equation (1)) to obtain a starting point for the shrink-wrap iterative reconstruction procedure. This iterative reconstruction procedure uses the $\chi^2$ error metric in Fourier space, as is typical for iterative CDI algorithms (see e.g. [40]):

$$\chi^2 = \sum_{i=1}^{N}\left(\sqrt{\hat{I}_{kin,i}} - \sqrt{Ir_i}\right)^2 \Big/ \sum_{i=1}^{N}\left(\hat{I}_{kin,i}\right). \tag{4}$$

In our case, $\hat{I}_{kin,i}$ is the 3D array of the simulated intensity distribution (with and without added noise) in Fourier space, modelled using the original values for the phase and amplitude. $Ir_i$ is the 3D array of the simulated intensity distribution (no extra noise added) in Fourier space, modelled using the reconstructed values for the phase and amplitude, produced by the iterative procedure.

To estimate errors in the reconstructed functions, with amplitude $\beta_h^{rec}(\mathbf{r})$, and phase $phase^{rec}(\mathbf{r})$, we use two metrics for the real-space data [42], namely a normalised root-mean-square (RMS) error criterion, defined as

$$d = \sqrt{\sum\left(G_{ijk}^{rec} - G_{ijk}^{ideal}\right)^2 \Big/ \sum\left(G_{ijk}^{ideal} - \langle G^{ideal}\rangle\right)^2}, \tag{5}$$

and a normalised absolute difference,

$$r = \sqrt{\sum\left|G_{ijk}^{rec} - G_{ijk}^{ideal}\right| \Big/ \sum\left|G_{ijk}^{ideal}\right|}, \tag{6}$$

where $G_{ijk}^{ideal}$ and $G_{ijk}^{rec}$ are ideal and reconstructed three-dimensional functions, respectively. $\langle G^{ideal}\rangle$ is the mean of the original function. As the phase reconstruction in the iterative procedure may contain an unknown constant offset, we calculate error metrics $d$ and $r$ for the derivative of the reconstructed phase, $phase^{rec}(\mathbf{r})$, with respect to the z coordinate, namely, $d\_ph\_z$ and $r\_ph\_z$, respectively. Tables 1 (no noise in the simulated data) and 2 (the maximum intensity of the simulated data is $10^{11}$ photons per voxel) show values of criteria $d\_ph\_z$, $r\_ph\_z$ and $d\_amp$, $r\_amp$ for the phase derivative along the z-direction ($\partial\left(phase^{rec}(\mathbf{r})\right)/\partial z$) and amplitude ($\beta_h^{rec}(\mathbf{r})$), respectively, for different reconstructions: deterministic BCDI, the shrink-wrap algorithm with deterministic BCDI as the starting point and the shrink-wrap algorithm with autocorrelation function as the starting point. For the (00L) type of reflections (i.e., $\mathbf{h}$ has only the z component) $\partial\left(phase^{rec}(\mathbf{r})\right)/\partial z$ is proportional to $\varepsilon_{zz} = \partial u_z/\partial z$ component of the symmetrical strain tensor $\varepsilon_{ij} = \tfrac{1}{2}\left(\partial u_j/\partial x_i + \partial u_i/\partial x_j\right)$ [43-45].

We restrict the total number of shrink-wrap iterations to 2000, because, as shown in Fig. 3, no improvements are observed beyond 1800 iterations. Figure 3 clearly demonstrates that the shrink-wrap iterative reconstruction procedure improves the $\chi^2$ metric for both starting points (deterministic BCDI or autocorrelation function) in real space in comparison to the one-step deterministic BCDI reconstruction, which is shown in Fig. 3 only for the case of the maximum intensity of $10^{11}$ photons per pixel, because it is indistinguishable from the noise-free reconstruction. The use of the deterministic BCDI reconstruction as the starting point in the iterative procedure allows faster convergence and the final $\chi^2$ result is better than the one obtained when the starting point is the autocorrelation function.

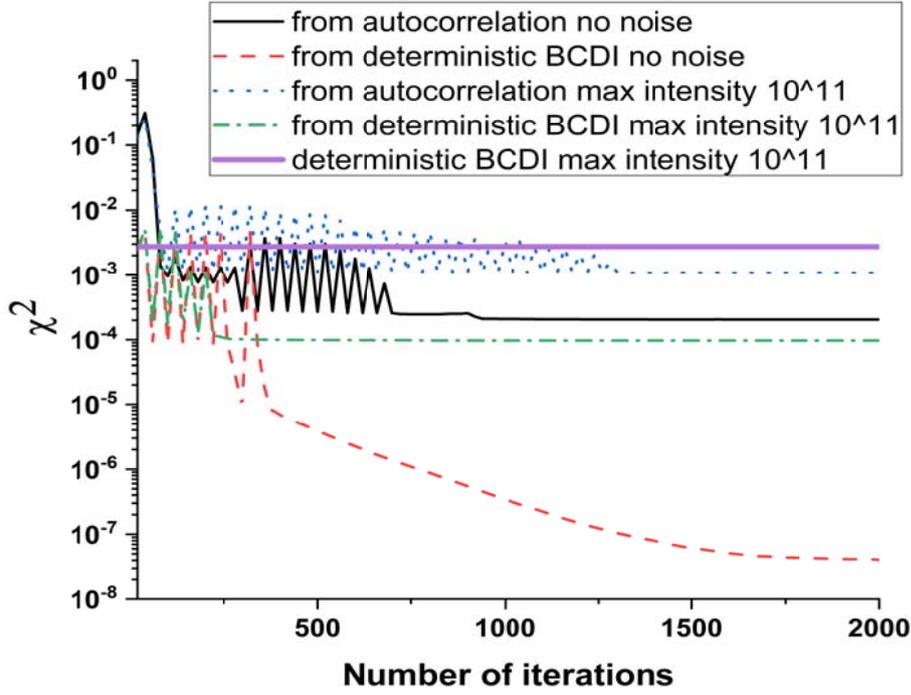

**Figure 3.** The $\chi^2$ error metrics as a function of the number of iterations.

In the case when no noise is added to the simulated intensity (equation (1)), the reconstruction results for the one-step deterministic BCDI reconstruction (($\beta_h^{rec}(\mathbf{r})$ and $phase^{rec}(\mathbf{r})$)) are shown in Fig. 2b.

Figure 2c shows real space outcomes ($\beta_h^{rec}(\mathbf{r})$ and $phase^{rec}(\mathbf{r})$) of the iterative procedure when the starting point was the deterministic BCDI reconstruction (Fig. 2b). The reconstruction errors, which are visible in Fig. 2b and are caused by the non-ideality of the reference part, have disappeared. At the same time the iterative procedure, using the autocorrelation function as its starting point, produces typical edge-like artefacts (see Fig. 2d) which have some artificial symmetry.

It should be noted that using only $\chi^2$ metrics does not give a proper indication of the reconstruction quality as demonstrated in Fig. 4. Certain error metrics can have a rapid drop or even a strong oscillation, while other error metrics stay relatively constant. For example, in Fig. 4(a), between 0 and 100 iterations, the red curve drops sharply and begins to oscillate, while the green curve drops only marginally and then stagnates; at around 700 iterations the roles are reversed, with the green metric dropping significantly while the red metric stays roughly constant. In Fig. 4(b), a steady drop of the red error metric over two orders of magnitude (from 300 to 1500 iterations) gives relatively little change in the pink curve; however, Fig. 4(d) shows the opposite behaviour, for at 1100 iterations both the blue and the pink error metrics drop significantly, while the black and red stay relatively constant, and the green metric even increases. No one error metric is able to give an unproblematic handle on quantifying all errors that may be present in the process of iterative refinement. A more balanced picture of the error landscape is provided by monitoring a portfolio of error metrics, as done in Figs 3 and 4. There can sometimes be relatively important changes in the reconstruction that are not accompanied by significant changes in the $\chi^2$ metric, corresponding to degrees of freedom for which a significant change in the properties of the scattering object yield only a small change in the corresponding CDI data.

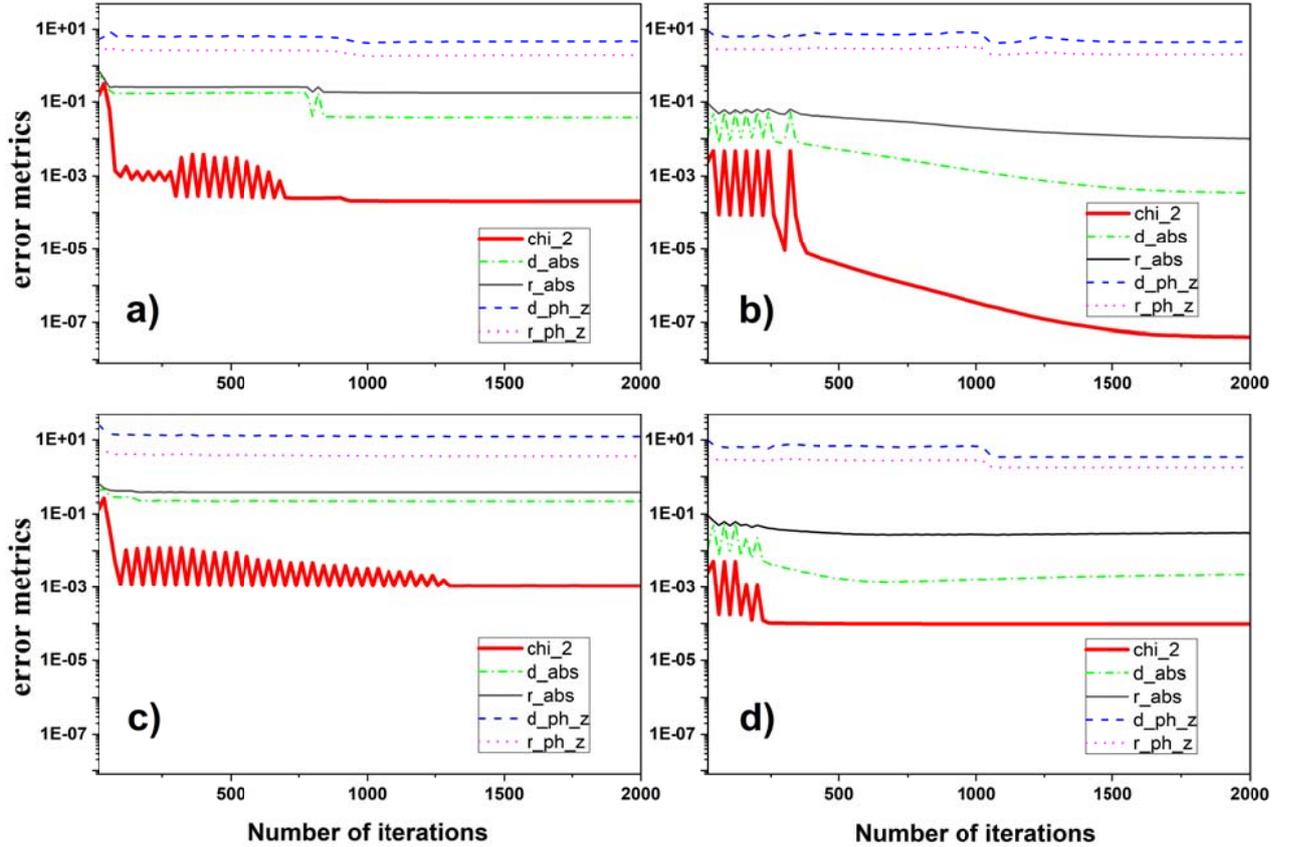

**Figure 4.** Real-space (d and r) and Fourier-space ($\chi^2$) error metrics as functions of the number of iterations. a) starting from autocorrelation, no noise; b) starting from DCDI, no noise; c) starting from autocorrelation, max intensity $10^{11}$ photons per voxel; d) starting from DCDI, max intensity $10^{11}$ photons per voxel. Real-space metrics (d_abs, r_abs) and (d_ph_z, r_ph_z) are applied for functions $\beta_h^{rec}(\mathbf{r})$ and $\partial\left(phase^{rec}(\mathbf{r})\right)/\partial z$, respectively.

## 4. Conclusions

In this simulations-based study we used a deterministic Bragg Coherent Diffraction Imaging approach to seed a subsequent shrink-wrap iterative refinement. The final reconstruction using this starting point was seen to be better than the corresponding approach that employs an autocorrelation function as a starting point, for both noisy and noise-free simulated data considered in this study. Two broad recommendations are suggested by this investigation. (i) Iterative CDI reconstructions may benefit from reconsidering the use of very crude first estimates, such as random starts or seeds based on auto-correlation estimates. The use of more refined estimates for subsequent iterative-CDI refinement, including but not limited to the deterministic CDI seed considered in this paper, may be useful for the further advancement of CDI reconstructions. (ii) Many iterative refinement schemes, including but not limited to those employed in CDI phase retrieval, monitor a single error metric to determine convergence. We suggest that a portfolio of such metrics, namely a vector of scalar error metrics, may provide a more nuanced measure with which to monitor convergence. Moreover, this vector of scalar error metrics gives additional degrees of freedom that may be input into each decision-tree node in iterative phase-retrieval algorithms.

**Acknowledgements**

KMP acknowledges financial support from the University of New England. VIP acknowledges financial support from the Russian Foundation for Basic Research (grants No 16-43-110350, No 17-02-00090) and the Ural branch of the Russian Academy of Sciences (grant No 18-10-2-23). KSM was supported by a Veski VPRF and completed this work with the support of the Technische Universität München Institute for Advanced Study, funded by the German Excellence Initiative and the European Union Seventh Framework Programme under grant agreement n° 291763 and co-funded by the European Union.

**Table 1** Error metrics r and d for real space and $\chi^2$ for Fourier space data for no-noise initial "experimental" intensity data. Real-space metrics (d_abs, r_abs) and (d_ph_z, r_ph_z) are for functions $\beta_h^{rec}(\mathbf{r})$ and $\partial\left(phase^{rec}(\mathbf{r})\right)/\partial z$, respectively. Non-perfect bottom half of crystal – a displacement field throughout the crystal. Maximum phase 0.25 rad. Minimum amplitude 0.9.

| | r_abs | r_ph_z | d_abs | d_ph_z | $\chi^2$ |
|---|---|---|---|---|---|
| DCDI | $1.43\times10^{-1}$ | $7.09\times10^{-1}$ | $4.23\times10^{-2}$ | $1.74\times10^{-1}$ | $2.74\times10^{-3}$ |
| iterations started from the DCDI data | $8.98\times10^{-3}$ | 2.17 | $3.27\times10^{-4}$ | 5.28 | $3.81\times10^{-8}$ |
| iterations started from the autocorrelation data | $3.74\times10^{-1}$ | 3.58 | $2.13\times10^{-1}$ | $1.23\times10^{1}$ | $1.08\times10^{-3}$ |

**Table 2** Error metrics r and d for real space and $\chi^2$ for Fourier space data for maximum "experimental" intensity of $10^{11}$ photons per voxel. Real-space metrics (d_abs, r_abs) and (d_ph_z, r_ph_z) are for functions $\beta_h^{rec}(\mathbf{r})$ and $\partial\left(phase^{rec}(\mathbf{r})\right)/\partial z$, respectively. Non-perfect bottom half of crystal – a displacement field throughout the crystal. Maximum phase 0.25 rad. Minimum amplitude 0.9.

| | r_abs | r_ph_z | d_abs | d_ph_z | $\chi^2$ |
|---|---|---|---|---|---|
| DCDI | $1.43\times10^{-1}$ | $7.09\times10^{-1}$ | $4.23\times10^{-2}$ | $1.74\times10^{-1}$ | $2.74\times10^{-3}$ |
| iterations started from the DCDI data | $3.13\times10^{-2}$ | 1.76 | $2.41\times10^{-3}$ | 3.42 | $9.69\times10^{-5}$ |
| iterations started from the autocorrelation data | $2.68\times10^{-1}$ | 1.95 | $2.50\times10^{-1}$ | 4.22 | $1.36\times10^{-4}$ |